\documentclass[cernpreprint, atlasdraft=false,coverpage=false , USenglish, texmf, orcidlogo]{atlasdoc}
\usepackage[subcaption]{atlaspackage}
\usepackage{atlasbiblatex}

\usepackage{atlasphysics}
\graphicspath{{logos/}{figures/}}

\AtlasTitle{Search for magnetic monopole pair production in ultraperipheral Pb+Pb collisions at  $\sqrt{s_{_\textrm{NN}}}=5.36$~TeV with the ATLAS detector at the LHC}

\AtlasAbstract{%
This Letter presents a search for highly ionizing magnetic monopoles in 262~$\mu$b$^{-1}$  of  ultraperipheral Pb+Pb collision data at $\sqrt{s_{_\textrm{NN}}}=5.36$~TeV collected by the ATLAS detector at the LHC. A new methodology that exploits the properties of clusters of hits reconstructed in the innermost silicon detector layers is introduced to study highly ionizing particles in heavy-ion data.
No significant excess above the background, which is estimated using a data-driven technique, is observed.
Using a nonperturbative semiclassical model, upper limits at 95\% confidence level  are set on the cross-section for pair production of monopoles with a single Dirac magnetic charge in the mass range of 20--150~GeV.
Depending on the model, monopoles with a single Dirac magnetic charge and mass below 80--120 GeV are excluded.
}

\AtlasRefCode{HION-2023-01}
\PreprintIdNumber{CERN-EP-2024-209}

\AtlasJournalRef{Phys. Rev. Lett. 134 (2025) 061803}
\AtlasDOI{DOI:10.1103/PhysRevLett.134.061803}

\hypersetup{pdftitle={ATLAS document},pdfauthor={The ATLAS Collaboration}}
\begin{document}
\newcommand{\xnxn}{\ensuremath{\mathrm{X}n\mathrm{X}n}}
\newcommand{\onxn}{\ensuremath{\mathrm{0}n\mathrm{X}n}}
\newcommand{\sqn}{\ensuremath{\sqrt{s_{_\text{NN}}}}\xspace}
\maketitle

%

%
%

%


%
%

Magnetic monopoles are hypothetical particles that carry isolated magnetic charge.
The existence of a magnetically charged particle would restore symmetry to Maxwell’s equations and explain why electric charge is quantized in nature, as shown by Dirac~\cite{Dirac:1931kp}.
Dirac's argument predicts the magnetic charge of a monopole to be $q_{\textrm{m}} = Ng_{\textrm{D}}$, where $N$ is a non-zero integer number, $g_{\textrm{D}} = e/(2\alpha) \approx 68.5e$ is the Dirac elementary charge in cgs Gaussian units, $\alpha$ is the fine-structure constant, and $e$ is the elementary electric charge.
Since the energy loss is proportional to the square of the charge, a monopole with $|q_{\textrm{m}}| = 1g_{\textrm{D}}$ would deposit 4700 times more energy by ionization than a proton. The high stopping power also results in the production of a large number of $\delta$-electrons.
Hence, magnetic monopoles would manifest as highly ionizing particles, with unique trajectories in a magnetic field~\cite{doi:10.1080/14786431003659180}.

While monopoles appearing in grand unification theories typically have masses of the order of the unification scale ($m \approx 10^{16}$~GeV)~\cite{HOOFT1974276} , some extensions of the Standard Model (SM) predict composite monopoles with masses near the TeV scale~\cite{Mavromatos:2020gwk}.
This improves the prospects for monopole production at existing colliders.

In ultrarelativistic heavy-ion collisions, the ion beams are accompanied by large electromagnetic (EM) fields.
At impact parameters $b$ larger than twice the nuclear radius, $b>2R_{\textrm{A}}$, EM-induced reactions become the dominant interaction mechanism.
These events, referred to as ultraperipheral collisions (UPC), have been used to study photon--nucleus (photonuclear) and photon--photon ($\gamma\gamma$) collisions~\cite{Klein:2020fmr}.
These events typically have features
such as rapidity gaps and lower particle multiplicity, or exclusive final states with no extra particle production, that make them qualitatively different from nuclear collisions where hadronic interactions occur~\cite{HION-2018-27}.
The interaction of strong magnetic fields in UPC can also give rise to the production of magnetic monopole-antimonopole pairs ($M\bar{M}$). Using a simplified (leading order, LO) approach, the process corresponds to a $\gamma\gamma$ fusion reaction, $\gamma\gamma\to M\bar{M}$~\cite{Epele:2012jn}.

Most searches for the direct production of magnetic monopoles at particle colliders have focused on collisions of elementary particles such as electrons or quarks, assuming pair production of spin-0 or spin-$\frac{1}{2}$ point-like monopoles via the Drell--Yan mechanism or $\gamma\gamma$ fusion process~\cite{PETRA:1983,PEP:1982,PEP:1984,CLEO:1987,TASSO:1988,TRISTAN:1988,TRISTAN:1989,OPAL:2008,MODAL:1992,L6-MODAL:1993,H1:2004,CDF:2005,Kalbfleisch:2003,Kalbfleisch:2000,D0:1998,Price:1987,Price:1990,Bertani:1990,Hoffmann:1978,Carrigan:1977,Aubert:1982,Carrigan:1973,Carrigan:1974,PhysRevLett.79.3134,EXOT-2011-27,EXOT-2014-16,EXOT-2017-20,EXOT-2019-33, MoEDAL2016a,MoEDAL2017a,MoEDAL2018,MoEDAL2019}.
However, because of the large coupling constant, $1/(4\alpha) \approx 34$~\cite{Song:2021vpo}, perturbative treatments in terms of Feynman diagrams are generally not well defined.
Indeed, it has been argued that the production of composite monopoles from elementary-particle collisions is suppressed by a factor of $\text{e}^{-4/\alpha} \approx 10^{-238}$~\cite{Drukier:1981fq}.

Magnetic monopoles could be produced in strong magnetic fields by a magnetic analog of the Schwinger mechanism~\cite{Schwinger:1951nm}.
The Schwinger effect describes how strong EM fields, particularly intense magnetic fields from ion collisions, can decay via quantum tunneling to generate particle pairs such as $M\bar{M}$ pairs.
In this case,
the $M\bar{M}$ production cross-section can be computed nonperturbatively using semiclassical techniques such as the free-particle approximation (FPA)~\cite{Gould:2019myj}.
In the FPA model the formula for the total cross-section for magnetic monopole production in Pb+Pb UPC~\cite{Gould:2019myj,Gould:2021bre,MoEDAL:2021vix} is
\begin{equation}
\sigma_{\textrm{FPA}} \approx \frac{2(q_{\textrm{m}}B)^{4}R_{\textrm{Pb}}^4}{9\pi^2m^5\omega} \exp{(-4m/\omega)}~,
\label{eq:sigma_fpa}
\end{equation}
where $m$ is the monopole mass, $q_{\textrm{m}}$ is the magnetic charge, $R_{\textrm{Pb}}=6.62\pm0.06$~fm is the radius of a lead nucleus, $B=(4.5\pm0.2)\cdot 10^{16}$~T is the peak magnetic field strength of the two nuclei at a nucleon--nucleon (NN) center-of-mass energy of $\sqn=5.36$~TeV, which occurs at $b\approx 2R_{\textrm{Pb}}$, and $\omega = (1.19\pm0.05)\cdot 10^{26}$~s$^{-1}$ is the field inverse decay time.
The uncertainties quoted on these parameters follow those estimated in Ref.~\cite{MoEDAL:2021vix}.
It should be noted that Eq.~(\ref{eq:sigma_fpa}) was derived for scalar (spin-0) monopoles, and the effect of monopole spin should enhance the production cross-section~\cite{Gould:2019myj}.
Because of the coherence of the magnetic field, the potential $\text{e}^{-4/\alpha}$ suppression is absent for composite monopole production via the Schwinger formalism in UPC.
This fact was used in the recent searches performed by the MoEDAL Collaboration in Pb+Pb collisions~\cite{MoEDAL:2021vix,MoEDAL:2024wbc} using the trapping technique~\cite{Milton:2006cp}, a method designed to physically capture magnetic monopoles and detect their presence by measuring the induced persistent currents generated after their passage through a superconducting magnetometer.

The FPA model includes the effect of the magnetic field to all orders but neglects possible $M\bar{M}$ self-interactions. Consequently, an alternative model, the locally constant field approximation (LCFA)~\cite{Gould:2019myj, Gould:2021bre, MoEDAL:2021vix, MoEDAL:2024wbc}, is also considered.
The LCFA includes self-interactions to all orders but neglects the effect of the nonperturbative interaction with the magnetic field.
Both approaches offer conservative lower limits on the production cross-section, as including the self-interactions and the effect of the magnetic field has been shown to increase the $M\bar{M}$ production~\cite{Gould:2019myj,Gould:2021bre}.
It is also worth mentioning that while the quoted semiclassical calculations apply to elementary monopoles, the effects of finite monopole size are expected to enhance monopole production in these models~\cite{Gould:2019myj}.

The mass of magnetic monopoles that could be produced in Pb+Pb collisions via the Schwinger mechanism is limited by the the total energy in the EM field available for $M\bar{M}$ production~\cite{MoEDAL:2024wbc}.
This total energy reaches its peak value at $b\approx 2R_{\textrm{Pb}}$.
For Pb+Pb collisions at the Large Hadron Collider (LHC) at $\sqn=5.36$~TeV, this implies a maximum monopole mass of approximately 160~GeV.

This Letter presents a search for
magnetic monopoles with $|q_{\textrm{m}}| = 1g_{\textrm{D}}$ using 262~$\mu$b$^{-1}$ of UPC Pb+Pb collision data collected by the ATLAS detector at the LHC in 2023 at $\sqn=5.36$~TeV.
The ATLAS experiment~\cite{PERF-2007-01, ATLAS:2023dns} is a multipurpose particle detector with cylindrical geometry~\cite{atlas-footnote}, comprising an inner detector~(ID) tracker surrounded by a thin superconducting solenoid, EM and hadronic calorimeters, and a muon spectrometer.
A software suite~\cite{SOFT-2022-02} is used in Monte Carlo (MC) simulation, in the reconstruction and analysis of real and simulated data, in detector operations, and in the trigger and data acquisition systems of the experiment.

This analysis primarily makes use of the ID pixel detector
and the zero-degree calorimeters~(ZDC).
The high-granularity silicon pixel detector surrounds the collision region.
The innermost pixel-detector layer, the IBL, consists of 280 pixel-sensor modules that cover the region $|\eta| < 3.03$, and was installed at a mean distance of 3.3~cm from the beam axis before the start of \RunTwo~\cite{ATLAS-TDR-2010-19,PIX-2018-001}.
In addition to the IBL, the pixel detector contains three other barrel layers and two endcaps with three disks each.
It is followed by silicon microstrip (SCT) and transition radiation tracking detectors.
ZDCs are located at $z = \pm$140~m from the interaction point, and detect neutral particles such as neutrons emitted from interacting nuclei.

\begin{figure}[!t]
\begin{center}
\subcaptionbox{\label{fig:sig_diagram_a}}{\includegraphics[width=0.3\textwidth]{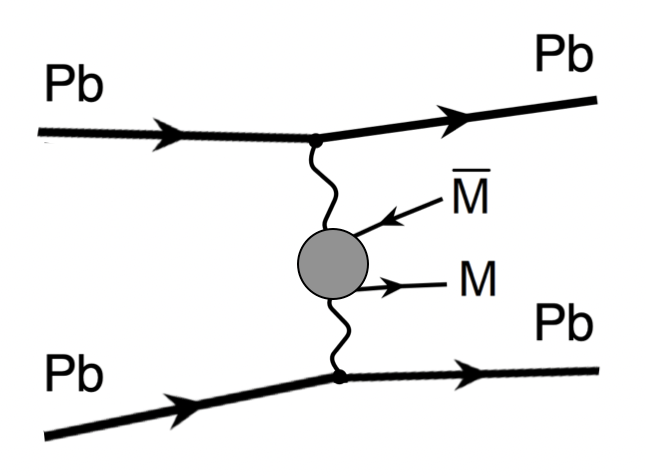}}
\subcaptionbox{\label{fig:sig_diagram_b}}{\includegraphics[width=0.3\textwidth]{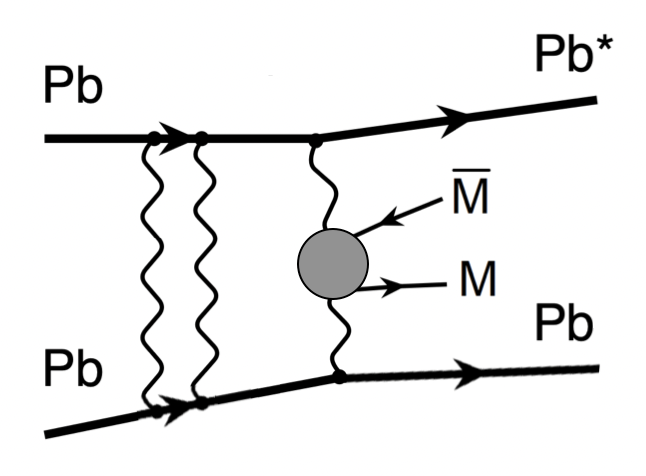}}
\subcaptionbox{\label{fig:sig_diagram_c}}{\includegraphics[width=0.3\textwidth]{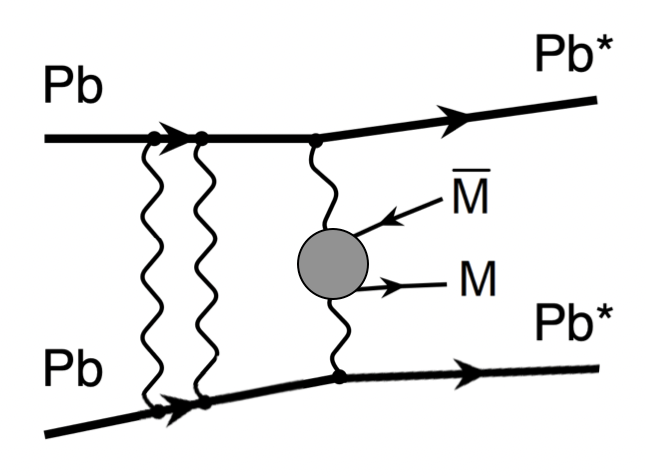}}
\caption{Schematic diagrams for magnetic monopole pair production in ultraperipheral Pb+Pb collisions for (a) no, (b) single, and (c) mutual ion excitation due to extra soft-photon exchanges. These three topologies are labeled in the text as 0$n$0$n$, 0$n$X$n$, and \xnxn, respectively. EM breakup of the ion, denoted by Pb$^*$, results in the production of forward neutrons, detectable in the ATLAS ZDC. The signal process in this study targets events with the \xnxn\ topology (c).}
\label{fig:sig_diagram}
\end{center}
\end{figure}

Events of interest are selected by the first-level (L1) trigger system implemented in custom hardware, followed by selections made by algorithms implemented in software in the high-level trigger (HLT)~\cite{ATLAS:2024xna}.
In UPC, soft photons emitted by one lead nucleus can excite the other, typically through the giant dipole resonance~\cite{Veyssiere:1970ztg}, and induce the emission of one or more neutrons, each of which carries, on average, the full per-nucleon beam energy.
This can give rise to three distinct EM breakup topologies: 0$n$0$n$ (no neutrons are emitted), 0$n$X$n$ (at least one neutron is emitted by exactly one nucleus) and \xnxn\ (at least one neutron is emitted by each nucleus).
They are shown schematically in Figure~\ref{fig:sig_diagram}.
This analysis exploits the \xnxn\ topology, mainly because of limitations in triggering on monopole production.
The ATLAS L1 trigger system can recognize energy deposited in the calorimeters, however, low-energy monopoles typically do not reach them. Moreover, no information from the ID tracker is available at L1.
Therefore, the only viable option is to trigger at L1 on the presence of forward neutrons in the ZDC.
The primary trigger for this analysis requires a L1 signal
consistent with the presence of one or more neutrons
in both arms of the ZDC.
In addition, the total transverse energy recorded in the calorimeter is required to be below 10~GeV.

As the monopoles considered in this analysis have relatively low masses (below 150~GeV) and typically have low energy, they deposit their energy in the innermost parts of the ATLAS detector, primarily in the pixel detector,  thereby producing a sizable number of $\delta$-electrons.
Hence this approach complements the trapping technique used by MoEDAL.
The presence of more than 100 clusters of pixel hits is required
in the HLT, with no specific selection on the presence of any charged-particle tracks.
Because of the large cross-section for double EM dissociation~\cite{ALICE:2012aa} and limitations in the allowed data-recording rate at L1, the trigger was prescaled so that approximately one event in every six was saved, resulting in an effective integrated luminosity of 262~$\mu$b$^{-1}$, compared to the 1.67~nb$^{-1}$ recorded by unprescaled triggers.
%


%

%


Monte Carlo simulated signal samples were produced using a semiclassical (nonperturbative) approach based on the FPA model.
The semiclassical approximation breaks down~\cite{MoEDAL:2021vix} for sufficiently light monopoles, $m<20$~GeV for 5.36~TeV Pb+Pb collisions, so monopole mass hypotheses below 20~GeV are not considered.
For the distributions of monopole kinematic quantities, simplified predictions based on the FPA model are used to model back-to-back monopole pair production, in particular with isotropic polar angle distributions of the pairs  and with a distribution of monopole momentum $k$ that follows the relative probability~\cite{Gould:2021bre}
\begin{equation*}
d \sigma_{\textrm{FPA}}(|k|)/d\sigma_{\textrm{FPA}}(0) = \exp{\left[-\frac{4}{\omega}\left(\sqrt{m^2+|k|^2}-m\right)\right]}~.
\label{momentum}
\end{equation*}
The same model was used in the MoEDAL analysis~\cite{MoEDAL:2021vix}.
A monopole charge of $|q_{\textrm{m}}| = 1g_\text{D}$ is set in all signal samples and the monopole masses considered are 20, 30, 40, 50, 60, 70, 90, 100, 120, and 150~GeV.
The signal simulation setup includes all possible configurations for the EM breakup of the nuclei, as described below.
A sample with no interacting particles was simulated to mimic detector noise and check for potential noise mismodeling, ensuring the reliability of the detector noise simulation in the signal modeling.

Signal efficiency estimates rely on an extension of the ATLAS detector simulation~\cite{SOFT-2010-01} based on \GEANT~\cite{Agostinelli:2002hh}.
This extension, originally developed for monopole searches in $pp$ collisions~\cite{EXOT-2011-27,EXOT-2014-16,EXOT-2017-20,EXOT-2019-33},
includes descriptions of monopole acceleration in the detector's magnetic field, ionization energy losses in matter, and $\delta$-electron production along the monopole's trajectory.
After losing their kinetic energy, monopoles could potentially get trapped in the detector material~\cite{Milton:2006cp}.
For all the simulated samples, the monopoles are assumed to be stable, and the simulation does not account for the trapping effect.


%

%


%
%

In order to correct the signal MC simulation for the \xnxn\ topology requirement, exclusive dilepton ($ee$ and $\mu\mu$) events from the process $\gamma\gamma\to\ell\ell$ are studied.
Because of the relatively large instantaneous luminosity of Pb+Pb collisions at the LHC, additional neutrons might be generated in each bunch crossing and detected in one or both arms of the ZDC (EM \pileup).
This leads to an outflow of events, primarily from the \onxn\ category, to the \xnxn\ category.
To account for this effect in MC simulation, the yield of simulated signal events is scaled by the effective probability for the \xnxn\ topology, parameterized as a function of total system mass $m_X$ and system rapidity $y_X$:
\begin{equation}
P_{\xnxn}^{\textrm{eff}}(m_X, y_X) = 0.8\times (2 f_{\onxn}\, P_{\textrm{EM}} + f_{\xnxn})(1+f_{\textrm{diss}})~,
\label{eq:peff}
\end{equation}
where $f_{\onxn}$ and $f_{\xnxn}$ are the predicted fractions of events with single and double EM breakup, respectively, $P_{\textrm{EM}}$ is the probability of having at least one neutron from EM \pileup on a given detector side, and $f_{\textrm{diss}}$ is the small fraction of dissociative (incoherent) events~\cite{diss} from the $\gamma^*\gamma\to \ell\ell$ and $\gamma^*\gamma^*\to \ell\ell$ reactions, where $\gamma^*$ denotes a virtual photon.
The $f_{\onxn}$ and $f_{\xnxn}$ values are based on the \textsc{SuperChic}\,4.2 MC predictions~\cite{Harland-Lang:2023ohq}, and are provided differentially in dilepton invariant mass ($m_{\ell\ell}$) and dilepton absolute rapidity ($|y_{\ell\ell}|$).
The value of $P_{\textrm{EM}}$, estimated as outlined in Ref.~\cite{HION-2016-02}, is $P_{\textrm{EM}} = 0.038$ for the signal trigger and 2023 Pb+Pb data-taking conditions.
The value of $f_{\textrm{diss}}$ for the \xnxn\ selection is estimated to be $f_{\textrm{diss}}=0.13$, based on the study of exclusive dimuon and dielectron events~\cite{HION-2016-02, HION-2021-16}.
A possible enhancement of the dissociative contribution at the largest considered monopole masses (due to very small Pb+Pb impact parameters), leading to larger values of $P_{\xnxn}^{\textrm{eff}}$, is conservatively neglected.
The factor of 0.8 in Eq.~(\ref{eq:peff}) accounts for mismodeling of $f_{\onxn}$ and $f_{\xnxn}$ in \textsc{SuperChic}, based on the calculation in Ref.~\cite{Harland-Lang:2023ohq}.
The model based on Eq.~(\ref{eq:peff}) is validated using UPC $\gamma\gamma\to\ell\ell$ data and good agreement (within 10\%) is found.
The value of $P_{\xnxn}^{\textrm{eff}}$ is about 15\% for $m_{X}=40$~GeV and grows to about 30\% for $m_{X}=200$~GeV.

The backgrounds considered in this analysis include those from particles produced in Pb+Pb interactions (collision events) and from beam-induced sources.
Beam-induced background (BIB) is caused by beam particle losses in the LHC ring upstream of the ATLAS experiment, due to interactions with residual gas within the beam pipe or with machine elements~\cite{DAPR-2012-01, DAPR-2014-01, DAPR-2017-01, DAPR-2021-02}.
It is characterized by particles moving almost parallel to the beam line.

A requirement of at most one reconstructed charged-particle track is used to suppress collision backgrounds and has only a minor impact on the signal efficiency because a monopole's trajectory in the axial magnetic field provided by a solenoid is straight in the $r{-}\phi$ plane and bends in the $r{-}z$ plane~\cite{DeRoeck:2011aa} (whereas the opposite applies to electrically charged particles).
Charged-particle tracks are required to pass the track selection, which is optimized to suppress combinatorial (fake) tracks in the dense track environment around Pb+Pb collisions~\cite{HION-2016-06}, and to have transverse momentum $p_\mathrm{T}>0.1$~\GeV, $|\eta|<2.5$, and a transverse impact parameter of $|d_0|<1$~mm calculated relative to the measured beam-line position.

Signal events are required to have at most one topological cluster of calorimeter-cell energy deposits~\cite{PERF-2014-07}, which effectively removes collision backgrounds with neutral particles.
These topoclusters, calibrated to take into account the noise suppression, must have transverse energy $E_\mathrm{T} > 0.1$~\GeV and $|\eta| < 4.9$.  They are also required to  meet the cell significance criteria for the measured energy as outlined in Ref.~\cite{STDM-2011-01} to suppress the contribution from electronic noise fluctuations.
Events are also required to have more than 150 pixel clusters~\cite{PERF-2012-05}, $n_{\textrm{PixCl}}> 150$, including more than 50 IBL clusters, $n_{\textrm{IBLCl}} > 50$.
Clusters from the four pixel-sensor modules observed to have abnormal noise distributions in data are not considered in the cluster counting.
Furthermore, events are rejected if the number of clusters in a particular module exceeds 90\% of all pixel clusters in the event.

The selection requirements imposed on tracks, topoclusters, and pixel clusters fully suppress the collision background.
In particular, the track and topocluster requirements are designed to remove hadronic Pb+Pb interactions, which typically involve a high multiplicity of tracks and larger energy deposits in the calorimeters. Furthermore, the additional requirement of a large number of pixel clusters specifically eliminates background from SM photon-induced processes in UPC, which tend to produce low hadronic activity and fewer pixel hits.

To further reduce the BIB, the azimuthal distribution of reconstructed pixel clusters is examined.
A variable $T$, inspired by the \textit{transverse thrust}~\cite{STDM-2017-28}, is defined as:
\begin{equation*}
T = (1/n_{\textrm{PixCl}}) \sum_{i=1}^{n_{\textrm{PixCl}}} |\hat{r}_{i} \cdot \hat{n}|~,
\label{eqn:1}
\end{equation*}
where $\hat{r}_{i}$ is the direction (unit vector) of a given pixel cluster in the transverse plane with respect to the origin of the ATLAS coordinate system, and the transverse direction $\hat{n}$ maximizes the expression and corresponds to an azimuthal angle $\phi_T$. The solution for $\hat{n}$ (or $\phi_T$)  is found iteratively. The direction of $\hat{n}$ roughly aligns in $r{-}\phi$ with the monopole's trajectory.
The $T$ variable has a maximum value of 1, for a set of fully aligned pixel clusters, and a minimum value of around $2/\pi$, for a uniform distribution of clusters in the transverse plane.
Simulated signal events tend to have $T$ values near unity, whereas the backgrounds
concentrate at much lower $T$ values, typically just above $2/\pi$.
A requirement of $T>0.95$ is therefore used in the signal region (SR) selection, based on the simulated signal properties.
The signal acceptance times efficiency of the SR selection varies between 4\% and 0.2\% for simulated events with low and high monopole masses, respectively, and is driven by the $n_{\textrm{PixCl}}> 150$ requirement.


%

%


%
%

The background yield in this analysis is estimated using a fully data-driven method.
A control region (CR1) is defined by requiring $T\leq 0.87$.
It is observed that the event characteristics in CR1 are consistent with those of BIB.
Since the BIB typically results in asymmetric event activity, an additional control region (CR2) enriched in BIB is defined from events passing a supporting trigger that
selects events with ZDC activity on one side and no activity on the opposite side.
The same event selection criteria as for the SR, except no requirement on $T$ and a different topocluster requirement, are applied to CR2.
To help enrich the CR2 sample in BIB events and consequently reduce the signal contamination, only events with 1--3
topoclusters are used, with at least one out-of-time energy deposit in the calorimeter, i.e., a topocluster with reconstructed time more than $10$~ns before the bunch-crossing time.
The presence of topoclusters with negative reconstructed time is one of the characteristic features of BIB~\cite{DAPR-2012-01}.
The signal contamination in both CR1 and CR2 is studied in the MC simulated samples and is found to be negligible.

\begin{figure}[t!]
\begin{center}
\subcaptionbox{\label{fig:sct_a}}{\includegraphics[width=0.47\textwidth]{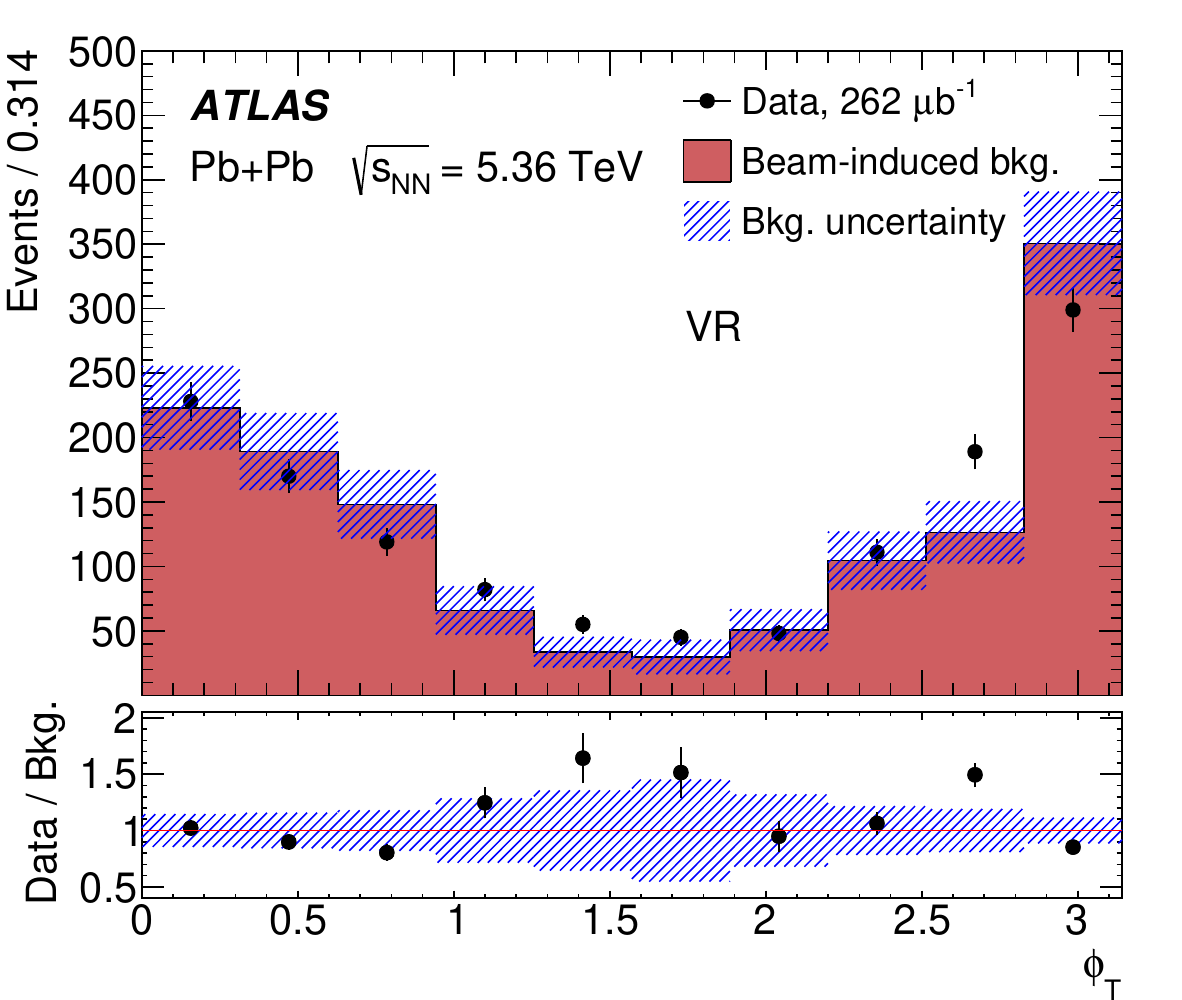}}
\subcaptionbox{\label{fig:sct_b}}{\includegraphics[width=0.47\textwidth]{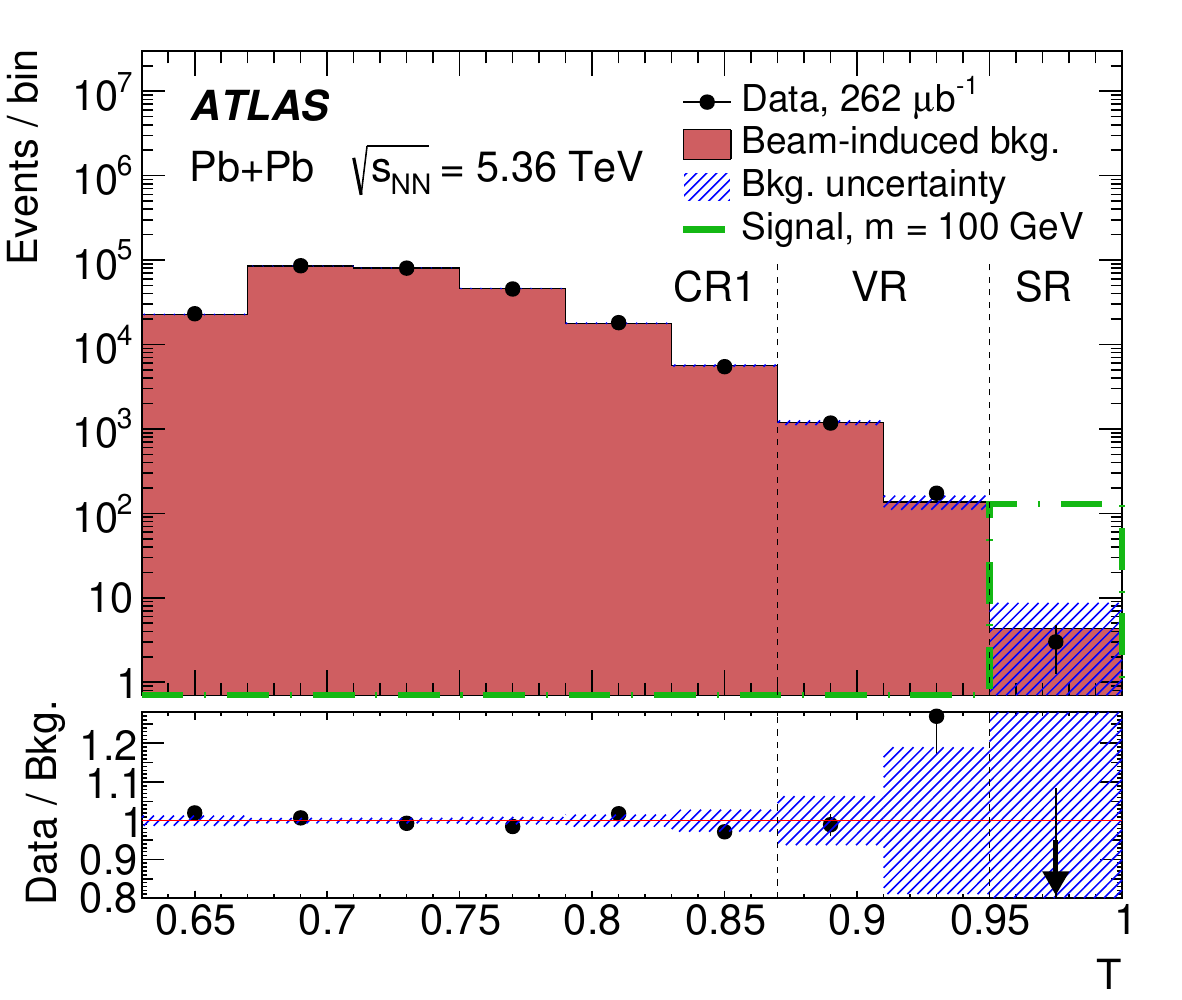}}
\caption{
(a) $\phi_T$ distribution for data in the VR and (b) $T$ distribution for data in CR1, the VR, and the SR.
Data (markers) are shown together with the estimated beam-induced background (filled histograms).
The distributions for background use events from CR2 scaled as described in the text. The lower panels show the ratio of data to the estimated background.
The shaded bands represent the statistical uncertainty of the background.
In (b) the green dotted-dashed line shows the representative signal contribution for a monopole mass of 100~\GeV, and the arrow in the ratio plot is for the point that is outside the range.}
\label{fig:sct}
\end{center}
\end{figure}

Events in CR2 are used to extrapolate the background contribution from CR1 to SR,
by using the relation: $N_{\textrm{bkg}}^{\textrm{SR}} = ({N^{\textrm{CR1}}/N_{T<0.87}^{\textrm{CR2}}})N_{T>0.95}^{\textrm{CR2}}$, where $N^{\textrm{CR1}}$ is the event yield in CR1, and $N_{T<0.87}^{\textrm{CR2}}$ ($N_{T>0.95}^{\textrm{CR2}}$) is the number of CR2 events having $T<0.87$ ($T>0.95$).
To cross-check this procedure, a validation region (VR) is defined by requiring $0.87< T\leq0.95$.
Figure~\ref{fig:sct_a} shows the
$\phi_T$ distributions in the VR.
The CR2-based background estimate describes the data adequately.
The enhanced event activity at $\phi_T\approx 0$ and $\phi_T\approx \pi$ is characteristic of BIB~\cite{DAPR-2012-01, DAPR-2014-01}.

The number of pixel clusters is correlated with the number of SCT space points~\cite{sct}, $n_{\text{SCTsp}}$. To address a small discrepancy between the CR1 and CR2 $n_{\text{SCTsp}}$ distributions at low $n_{\text{SCTsp}}$, extra reweighting of the $n_{\text{SCTsp}}$ distribution is performed for events in CR2 to better match the distribution observed in CR1. The correction applied to the $n_{\text{SCTsp}}$ distribution improves the modeling of the $n_{\textrm{PixCl}}$ distribution. This correction changes the shape of the $T$ distribution by a few percent at low $T$ and by about 10\% at high $T$.

Figure~\ref{fig:sct_b} shows the $T$ distribution in CR1, the VR, and the SR.
The background in the SR is estimated to be $4 \pm 4$~(stat.) events.

%
%
%
%
%
%
%
%
%
%
%
%
%


%

%


%
%
The systematic uncertainties that were considered are related to the modeling of the detector response to the monopole, the overall noise level in the pixel detector, potential mismodeling of the \xnxn\ selection, the background uncertainty, and the luminosity uncertainty. Uncertainties evaluated in one direction are assumed to be symmetric.

The uncertainty due to the ID material modeling in the \GEANT simulation is accounted for by comparing the signal efficiency with its value in alternative signal samples.
These samples differ by having modified descriptions of the ATLAS ID geometry:
the passive material of the ID is scaled up by 5\%, the passive material of the IBL is scaled up by 10\%, or the passive material in the services region is scaled up by 25\%.
These variations capture the full range of data--MC differences observed in studies of the ID material~\cite{PERF-2015-07}.
The changes in signal yield depend on the monopole mass hypothesis and range from 3\% to 20\% for the lowest and highest mass points, respectively.

The kinetic energy threshold below which $\delta$-electrons are not propagated explicitly in the ATLAS simulation depends on the \GEANT \enquote{range cut} parameter~\cite{Rimoldi:1097789}. Reducing the value of this parameter by a factor of five produces less than a 3\% signal yield change, which is taken as a systematic uncertainty.

The modified Bethe--Bloch formula for ionization by monopoles has a theoretical uncertainty of about $\pm$3\%~\cite{Ahlen:1980xr}.
Alternative signal samples where $\delta$-electron production is reduced by 3\% were therefore simulated. This results in a 1\%--4\% decrease in signal yield for the mass range considered, which is included as a systematic uncertainty.

Noise activity in the pixel detector is not fully accounted for in the simulation.
Less activity is observed in MC simulated events with no interacting particles than in \enquote{empty} data events with no tracks or topoclusters, and with at least one neutron emitted by the EM breakup of an ion.
These data events, selected by an unbiased trigger that requires ZDC activity on one side, are fully equivalent to empty bunch-crossing events.
To obtain better data--MC agreement, the MC simulated \enquote{empty} events are overlayed, on an event-by-event basis, with additional pixel clusters randomly assigned to pixel-cluster positions seen in \enquote{empty} data events.
The same degree of pixel-cluster overlay is then applied in MC signal events, resulting in a 1\%--2\% decrease in signal yield, which is taken as a systematic uncertainty.
A similar procedure is used to estimate calorimeter noise, resulting in a 1\% decrease in signal yield.

To cover the \xnxn\ selection modeling, a 20\% uncertainty is assigned to $p_{\xnxn}^{\textrm{eff}}$.
A value of 20\% is chosen to primarily cover the 10\%--20\% differences observed between the $f_{\onxn}$ and $f_{\xnxn}$ values in ATLAS data and those predicted by \textsc{SuperChic}\,4.2~\cite{Harland-Lang:2023ohq}.
It also covers differences between \textsc{SuperChic} and alternative \textsc{STARlight}~\cite{Klein:2016yzr} or \textsc{gamma-UPC}~\cite{Shao:2022cly} predictions for $f_{\onxn}$ and $f_{\xnxn}$.

The difference between the background estimates from the reweighted and non-reweighted $n_{\text{SCTsp}}$ distributions in CR2 affects the background yield in SR by 10\% and is taken as a systematic uncertainty.
The reweighting of other distributions that show slight differences between CR1 and CR2 has a negligible impact on the background estimate.

The uncertainty in the integrated luminosity of the data sample is 3.5\%.
It is derived from the calibration of the luminosity following a methodology similar to that detailed in Ref.~\cite{DAPR-2021-01}, and using the LUCID-2 detector~\cite{LUCID2} for the baseline luminosity measurements. The total systematic uncertainty affecting the selection efficiency, computed by adding in quadrature all of the individual systematic uncertainties, varies between 21\% (lowest masses) and 38\% (highest masses).


%

%


%
%

\begin{figure}[t!]
\begin{center}
\includegraphics[width=0.62\textwidth]{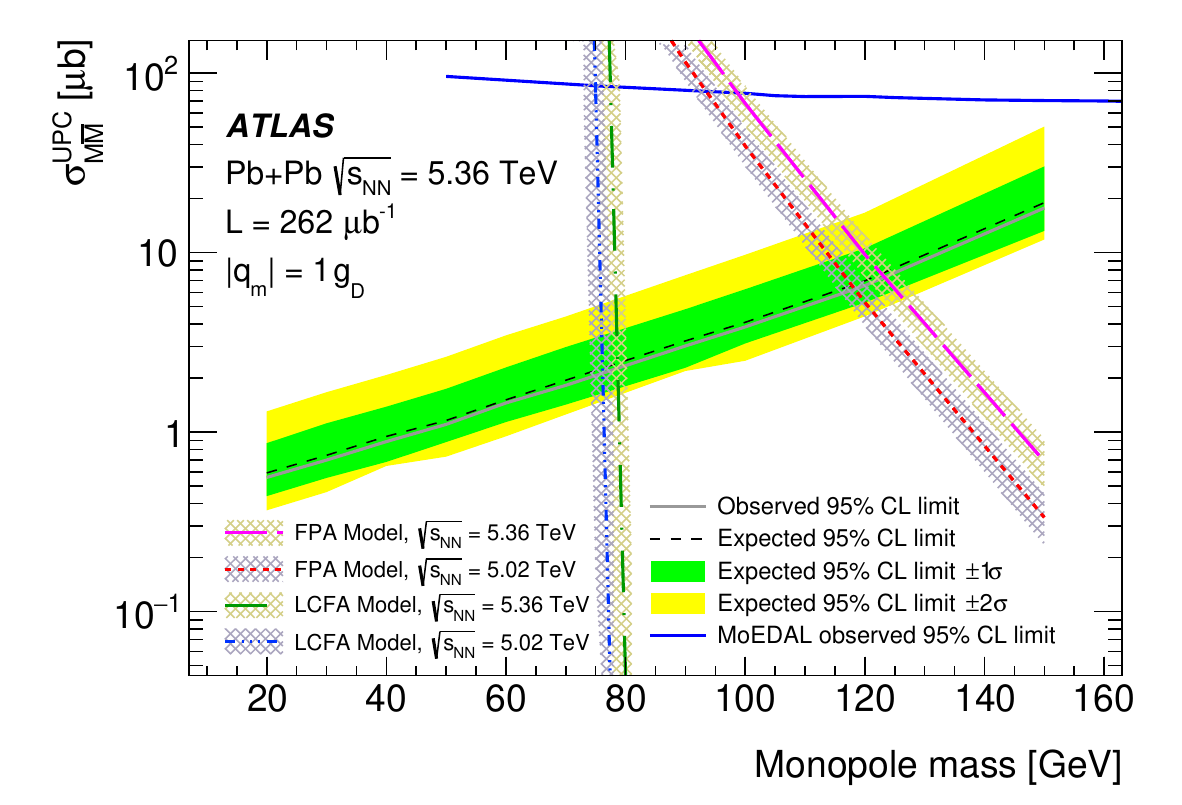}
\end{center}
\caption{Expected and observed upper limits on the monopole pair-production cross-section in Pb+Pb UPC at $\sqn=5.36$~TeV for $|q_{\textrm{m}}|=1g_{\textrm{D}}$. The gray solid line (black dashed line) represents observed (expected) limits, whereas the darker and lighter shaded bands around the expected limits represent the $\pm$1$\sigma$ and $\pm$2$\sigma$ intervals, respectively.
The limits are compared with the observed limits by MoEDAL for $|q_{\textrm{m}}|=1g_{\textrm{D}}$ at a lower center-of-mass energy of $\sqn=5.02$~TeV~\cite{MoEDAL:2021vix} (blue line), and with two model predictions: FPA and LCFA~\cite{Gould:2019myj}.
The two sets of model curves (5.36 TeV and 5.02 TeV) are shown to allow a comparison of the theoretical predictions at different center-of-mass energies, relevant for ATLAS and MoEDAL experiments, respectively.
The hatched bands represent the uncertainty of the models, estimated by propagating the uncertainties of the model parameters.
}
\label{fig:limits}
\end{figure}

Three data events were found in the SR. This is consistent with the estimate of $4 \pm 4\,\text{(stat.)} \pm 1\,\text{(syst.)}$ background events.
Consequently, exclusion limits are set at 95\% confidence level (CL) using the CL$_\text{s}$ method~\cite{Read:2002hq} implemented in RooStats~\cite{Cranmer:1456844}.
The cross-section limits are obtained by exploiting the selection efficiency and its uncertainty for each signal sample, the systematic uncertainty of the background estimate, and the integrated luminosity's uncertainty.
Limits are calculated using the pseudo-experiment approach, with 20\;\!000 \enquote{toys} per mass point.
Figure~\ref{fig:limits} shows the obtained 95\% CL limits.
These results are compared with the observed limits by MoEDAL~\cite{MoEDAL:2021vix}.
In addition, a comparison is made with the FPA model predictions~\cite{Gould:2019myj} (Eq.~(\ref{eq:sigma_fpa})) and with the alternative LCFA model~\cite{Gould:2019myj, MoEDAL:2021vix, MoEDAL:2024wbc}.
The present search significantly improves on the previous cross-section limits reported by MoEDAL for $|q_{\textrm{m}}|=1g_{\textrm{D}}$ and excludes monopoles with masses below 120~GeV, assuming the FPA model.
Assuming the total cross-section follows the LCFA model, magnetic monopoles with charge $|q_{\textrm{m}}|=1g_{\textrm{D}}$ and masses below 80~GeV are excluded.
Limits obtained from alternative production channels, like from indirect searches of monopoles produced close to the surface of neutron stars via the thermal Schwinger process~\cite{Gould:2017zwi}, are much less stringent and are not shown here.
%


%

%

%
%
%

%
%
%
%
%
%
%
%
%

In conclusion, this Letter presents a search for magnetic monopoles with mass in the range 20--150~GeV in ultraperipheral heavy-ion collisions using 262~$\mu$b$^{-1}$ of Pb+Pb collision data at $\sqn=5.36$~TeV collected by the ATLAS detector at the LHC.
This analysis uses an alternative way of detecting low-mass monopoles in heavy-ion collisions, complementary to the trapping technique used by the MoEDAL experiment.
The targeted monopole signature is based on high ionization in the ATLAS pixel detector.
The background is mainly beam-induced, and its yield is estimated using a data-driven procedure.
No excess of events over the expected background is observed.
For monopoles with a single Dirac magnetic charge, the derived upper limits on monopole pair-production cross-sections, based on a non-perturbative semiclassical model and derived at 95\% confidence level, are more stringent than the recently reported limits from MoEDAL, also using Pb+Pb collisions.
Depending on the model, monopoles with a single Dirac magnetic charge and mass below 80--120 GeV are excluded.
The exploration of higher magnetic charges will be a subject of future investigations.


%

%

%
\section*{Acknowledgments}
%

%
%

%
%

We thank CERN for the very successful operation of the LHC and its injectors, as well as the support staff at
CERN and at our institutions worldwide without whom ATLAS could not be operated efficiently.

The crucial computing support from all WLCG partners is acknowledged gratefully, in particular from CERN, the ATLAS Tier-1 facilities at TRIUMF/SFU (Canada), NDGF (Denmark, Norway, Sweden), CC-IN2P3 (France), KIT/GridKA (Germany), INFN-CNAF (Italy), NL-T1 (Netherlands), PIC (Spain), RAL (UK) and BNL (USA), the Tier-2 facilities worldwide and large non-WLCG resource providers. Major contributors of computing resources are listed in Ref.~\cite{ATL-SOFT-PUB-2023-001}.

We gratefully acknowledge the support of ANPCyT, Argentina; YerPhI, Armenia; ARC, Australia; BMWFW and FWF, Austria; ANAS, Azerbaijan; CNPq and FAPESP, Brazil; NSERC, NRC and CFI, Canada; CERN; ANID, Chile; CAS, MOST and NSFC, China; Minciencias, Colombia; MEYS CR, Czech Republic; DNRF and DNSRC, Denmark; IN2P3-CNRS and CEA-DRF/IRFU, France; SRNSFG, Georgia; BMBF, HGF and MPG, Germany; GSRI, Greece; RGC and Hong Kong SAR, China; ISF and Benoziyo Center, Israel; INFN, Italy; MEXT and JSPS, Japan; CNRST, Morocco; NWO, Netherlands; RCN, Norway; MNiSW, Poland; FCT, Portugal; MNE/IFA, Romania; MSTDI, Serbia; MSSR, Slovakia; ARIS and MVZI, Slovenia; DSI/NRF, South Africa; MICIU/AEI, Spain; SRC and Wallenberg Foundation, Sweden; SERI, SNSF and Cantons of Bern and Geneva, Switzerland; NSTC, Taipei; TENMAK, T\"urkiye; STFC/UKRI, United Kingdom; DOE and NSF, United States of America.

Individual groups and members have received support from BCKDF, CANARIE, CRC and DRAC, Canada; CERN-CZ, FORTE and PRIMUS, Czech Republic; COST, ERC, ERDF, Horizon 2020, ICSC-NextGenerationEU and Marie Sk{\l}odowska-Curie Actions, European Union; Investissements d'Avenir Labex, Investissements d'Avenir Idex and ANR, France; DFG and AvH Foundation, Germany; Herakleitos, Thales and Aristeia programmes co-financed by EU-ESF and the Greek NSRF, Greece; BSF-NSF and MINERVA, Israel; NCN and NAWA, Poland; La Caixa Banking Foundation, CERCA Programme Generalitat de Catalunya and PROMETEO and GenT Programmes Generalitat Valenciana, Spain; G\"{o}ran Gustafssons Stiftelse, Sweden; The Royal Society and Leverhulme Trust, United Kingdom.

In addition, individual members wish to acknowledge support from Armenia: Yerevan Physics Institute (FAPERJ); CERN: European Organization for Nuclear Research (CERN PJAS); Chile: Agencia Nacional de Investigaci\'on y Desarrollo (FONDECYT 1230812, FONDECYT 1230987, FONDECYT 1240864); China: Chinese Ministry of Science and Technology (MOST-2023YFA1605700, MOST-2023YFA1609300), National Natural Science Foundation of China (NSFC - 12175119, NSFC 12275265, NSFC-12075060); Czech Republic: Czech Science Foundation (GACR - 24-11373S), Ministry of Education Youth and Sports (FORTE CZ.02.01.01/00/22\_008/0004632), PRIMUS Research Programme (PRIMUS/21/SCI/017); EU: H2020 European Research Council (ERC - 101002463); European Union: European Research Council (ERC - 948254, ERC 101089007), Horizon 2020 Framework Programme (MUCCA - CHIST-ERA-19-XAI-00), European Union, Future Artificial Intelligence Research (FAIR-NextGenerationEU PE00000013), Italian Center for High Performance Computing, Big Data and Quantum Computing (ICSC, NextGenerationEU); France: Agence Nationale de la Recherche (ANR-20-CE31-0013, ANR-21-CE31-0013, ANR-21-CE31-0022, ANR-22-EDIR-0002), Investissements d'Avenir Labex (ANR-11-LABX-0012); Germany: Baden-Württemberg Stiftung (BW Stiftung-Postdoc Eliteprogramme), Deutsche Forschungsgemeinschaft (DFG - 469666862, DFG - CR 312/5-2); Italy: Istituto Nazionale di Fisica Nucleare (ICSC, NextGenerationEU), Ministero dell'Università e della Ricerca (PRIN - 20223N7F8K - PNRR M4.C2.1.1); Japan: Japan Society for the Promotion of Science (JSPS KAKENHI JP22H01227, JSPS KAKENHI JP22H04944, JSPS KAKENHI JP22KK0227, JSPS KAKENHI JP23KK0245); Netherlands: Netherlands Organisation for Scientific Research (NWO Veni 2020 - VI.Veni.202.179); Norway: Research Council of Norway (RCN-314472); Poland: Ministry of Science and Higher Education (IDUB AGH, POB8, D4 no 9722), Polish National Agency for Academic Exchange (PPN/PPO/2020/1/00002/U/00001), Polish National Science Centre (NCN 2021/42/E/ST2/00350, NCN OPUS nr 2022/47/B/ST2/03059, NCN UMO-2019/34/E/ST2/00393, NCN \& H2020 MSCA 945339, UMO-2020/37/B/ST2/01043, UMO-2021/40/C/ST2/00187, UMO-2022/47/O/ST2/00148, UMO-2023/49/B/ST2/04085, UMO-2023/51/B/ST2/00920); Slovenia: Slovenian Research Agency (ARIS grant J1-3010); Spain: Generalitat Valenciana (Artemisa, FEDER, IDIFEDER/2018/048), Ministry of Science and Innovation (MCIN \& NextGenEU PCI2022-135018-2, MICIN \& FEDER PID2021-125273NB, RYC2019-028510-I, RYC2020-030254-I, RYC2021-031273-I, RYC2022-038164-I); Sweden: Carl Trygger Foundation (Carl Trygger Foundation CTS 22:2312), Swedish Research Council (Swedish Research Council 2023-04654, VR 2018-00482, VR 2022-03845, VR 2022-04683, VR 2023-03403, VR grant 2021-03651), Knut and Alice Wallenberg Foundation (KAW 2018.0458, KAW 2019.0447, KAW 2022.0358); Switzerland: Swiss National Science Foundation (SNSF - PCEFP2\_194658); United Kingdom: Leverhulme Trust (Leverhulme Trust RPG-2020-004), Royal Society (NIF-R1-231091); United States of America: U.S. Department of Energy (ECA DE-AC02-76SF00515), Neubauer Family Foundation.

%
%


%
%
%
%
%
%

%
%

%
%
%
\printbibliography

@article{DeRoeck:2011aa,
    author = "De Roeck, A. and Katre, A. and Mermod, P. and Milstead, D. and Sloan, T.",
    title = "{Sensitivity of LHC Experiments to Exotic Highly Ionising Particles}",
    eprint = "1112.2999",
    archivePrefix = "arXiv",
    primaryClass = "hep-ph",
    doi = "10.1140/epjc/s10052-012-1985-2",
    journal = "Eur. Phys. J. C",
    volume = "72",
    pages = "1985",
    year = "2012"
}

@article{doi:10.1080/14786431003659180,
author = {Maxwell, J. C.},
title = {On physical lines of force},
journal = {Phil. Mag.},
volume = {90},
pages = {11},
year = {1861},
doi = {10.1080/14786446108643033},

}

@article{Milton:2006cp,
    author = "Milton, Kimball A.",
    title = "{Theoretical and experimental status of magnetic monopoles}",
    eprint = "hep-ex/0602040",
    archivePrefix = "arXiv",
    doi = "10.1088/0034-4885/69/6/R02",
    journal = "Rept. Prog. Phys.",
    volume = "69",
    pages = "1637--1712",
    year = "2006"
}

@misc{atlas-footnote,
  Note = {ATLAS uses a right-handed coordinate system with its origin at the nominal interaction point (IP) in the center of the detector and the $z$ axis coinciding with the axis of the beam pipe. The $x$ axis points from the IP to the center of the LHC ring, and the $y$ axis points upward. Cylindrical coordinates $(r, \phi)$ are used in the transverse plane, $\phi$ being the azimuthal angle around the beam pipe. The pseudorapidity is defined in terms of the polar angle $\theta$ as $\eta =-\ln \tan(\theta/2)$. } 
}

@misc{sct,
  Note = {An SCT space point is a pair of matched clusters on two sides of an SCT detector module } 
}

@misc{diss,
  Note = {The omission of the dissociative contribution to the signal, motivated by the possible suppression of monopole-pair production in the elementary $\gamma\gamma$ fusion process in semiclassical models,  weakens the cross-section limits by about 10\%. } 
}

@article{ATLAS:2024xna,
    author = "{ATLAS Collaboration}",
    collaboration = "ATLAS",
    title = "{The ATLAS Trigger System for LHC Run 3 and Trigger performance in 2022}",
    eprint = "2401.06630",
    archivePrefix = "arXiv",
    primaryClass = "hep-ex",
    reportNumber = "CERN-EP-2023-299",
    doi = "10.1088/1748-0221/19/06/P06029",
    journal = "JINST",
    volume = "19",
    pages = "P06029",
    year = "2024"
}

@article{ATLAS:2023dns,
    author = "{ATLAS Collaboration}",
    collaboration = "ATLAS",
    title = "{The ATLAS experiment at the CERN Large Hadron Collider: a description of the detector configuration for Run~3}",
    eprint = "2305.16623",
    archivePrefix = "arXiv",
    primaryClass = "physics.ins-det",
    reportNumber = "CERN-EP-2022-259",
    doi = "10.1088/1748-0221/19/05/P05063",
    journal = "JINST",
    volume = "19",
    number = "05",
    pages = "P05063",
    year = "2024"
}

@article{Shao:2022cly,
    author = "Shao, Hua-Sheng and d'Enterria, David",
    title = "{gamma-UPC: automated generation of exclusive photon-photon processes in ultraperipheral proton and nuclear collisions with varying form factors}",
    eprint = "2207.03012",
    archivePrefix = "arXiv",
    primaryClass = "hep-ph",
    doi = "10.1007/JHEP09(2022)248",
    journal = "JHEP",
    volume = "09",
    pages = "248",
    year = "2022"
}

@article{DAPR-2021-02,
    author = "{ATLAS Collaboration}",
    collaboration = "ATLAS",
    title = "{Beam-induced backgrounds measured in the ATLAS detector during local gas injection into the LHC beam vacuum}",
    eprint = "2405.05054",
    archivePrefix = "arXiv",
    primaryClass = "physics.ins-det",
    reportNumber = "CERN-EP-2024-113",
    doi = "10.1088/1748-0221/19/06/P06014",
    journal = "JINST",
    volume = "19",
    number = "06",
    pages = "P06014",
    year = "2024"
}

@article{Epele:2012jn,
    author = "Epele, Luis N. and Fanchiotti, Huner and Garc\'ia Canal, Carlos A. and Mitsou, Vasiliki A. and Vento, Vicente",
    title = "{Looking for magnetic monopoles at LHC with diphoton events}",
    eprint = "1205.6120",
    archivePrefix = "arXiv",
    primaryClass = "hep-ph",
    doi = "10.1140/epjp/i2012-12060-8",
    journal = "Eur. Phys. J. Plus",
    volume = "127",
    pages = "60",
    year = "2012"
}

@article{Gould:2017zwi,
    author = "Gould, Oliver and Rajantie, Arttu",
    title = "{Magnetic monopole mass bounds from heavy ion collisions and neutron stars}",
    eprint = "1705.07052",
    archivePrefix = "arXiv",
    primaryClass = "hep-ph",
    reportNumber = "IMPERIAL-TP-2017-OG-2",
    doi = "10.1103/PhysRevLett.119.241601",
    journal = "Phys. Rev. Lett.",
    volume = "119",
    number = "24",
    pages = "241601",
    year = "2017"
}

@article{ALICE:2012aa,
    author = "{ALICE Collaboration}",
    collaboration = "ALICE",
    title = "{Measurement of the Cross Section for Electromagnetic Dissociation with Neutron Emission in Pb-Pb Collisions at $\sqrt{s_{NN}}$ = 2.76 TeV}",
    eprint = "1203.2436",
    archivePrefix = "arXiv",
    primaryClass = "nucl-ex",
    reportNumber = "CERN-PH-EP-2012-065",
    doi = "10.1103/PhysRevLett.109.252302",
    journal = "Phys. Rev. Lett.",
    volume = "109",
    pages = "252302",
    year = "2012"
}

@Booklet{Cranmer:1456844,
      author        = "Cranmer, Kyle and Lewis, George and Moneta, Lorenzo and
                       Shibata, Akira and Verkerke, Wouter",
      title         = "{HistFactory: A tool for creating statistical models for
                       use with RooFit and RooStats}",
      collaboration = "ROOT Collaboration",
      howpublished  = "{CERN-OPEN-2012-016}",
      month         = "1",
      year          = "2012",
      url           = "https://cds.cern.ch/record/1456844",
}

@article{LUCID2,
  author={G. Avoni and others},
  title={The new LUCID-2 detector for luminosity measurement and monitoring in ATLAS},
  journal={JINST},
  volume={13},
  number={07},
  pages={P07017},
  doi="10.1088/1748-0221/13/07/P07017",
  year={2018},
}

@article{Schwinger:1951nm,
    author = "Schwinger, Julian",
    title = "{On gauge invariance and vacuum polarization}",
    doi = "10.1103/PhysRev.82.664",
    journal = "Phys. Rev.",
    volume = "82",
    pages = "664--679",
    year = "1951"
}

@article{Veyssiere:1970ztg,
    author = "Veyssiere, A. and Beil, H. and Bergere, R. and Carlos, P. and Lepretre, A.",
    title = "{Photoneutron cross sections of 208 Pb and 197 Au}",
    doi = "10.1016/0375-9474(70)90727-X",
    journal = "Nucl. Phys. A",
    volume = "159",
    pages = "561--576",
    year = "1970"
}

@article{Song:2021vpo,
    author = "Song, W.-Y. and Taylor, Wendy",
    title = "{Pair production of magnetic monopoles and stable high-electric-charge objects in proton\textendash{}proton and heavy-ion collisions}",
    eprint = "2107.10789",
    archivePrefix = "arXiv",
    primaryClass = "hep-ph",
    doi = "10.1088/1361-6471/ac3dce",
    journal = "J. Phys. G",
    volume = "49",
    number = "4",
    pages = "045002",
    year = "2022"
}

@article{Dirac:1931kp,
    author = "Dirac, Paul Adrien Maurice",
    title = "{Quantised singularities in the electromagnetic field,}",
    reportNumber = "RX-722",
    doi = "10.1098/rspa.1931.0130",
    journal = "Proc. Roy. Soc. Lond. A",
    volume = "133",
    number = "821",
    pages = "60--72",
    year = "1931"
}

@article{PETRA:1983,
    author = "Musset, P. and Price, M. and Lohrmann, E.",
    title = "{Search for magnetic monopoles in electron-positron collisions at 34 GeV CM energy}",
    reportNumber = "PRINT-83-0602 (HAMBURG), DESY-83-053",
    doi = "10.1016/0370-2693(83)90270-8",
    journal = "Phys. Lett. B",
    volume = "128",
    pages = "333",
    year = "1983"
}

@article{PEP:1982,
    author = "Kinoshita, K. and Price, P. B. and Fryberger, D.",
    title = "{Search for Highly Ionizing Particles in ${e}^{+}{e}^{\ensuremath{-}}$ Collisions at $\sqrt{s}=29$ GeV}",
    doi = "10.1103/PhysRevLett.48.77",
    journal = "Phys. Rev. Lett.",
    volume = "48",
    pages = "77",
    year = "1982"
}

@article{PEP:1984,
    author = "Fryberger, D. and Coan, T. E. and Kinoshita, K. and Price, P. B.",
    title = "{Search for highly ionizing particles in $e^+ e^-$ collisions at $\sqrt{s}=29$ {GeV}}",
    doi = "10.1103/PhysRevD.29.1524",
    journal = "Phys. Rev. D",
    volume = "29",
    pages = "1524",
    year = "1984"
}

@article{CLEO:1987,
    author = "Gentile, T. and others", 
    %collaboration = "CLEO Collaboration",
    title = "{Search for magnetically charged particles produced in $e^+ e^-$ annihilations at $\sqrt{s}=10.6$~GeV}",
    doi = "10.1103/PhysRevD.35.1081",
    journal = "Phys. Rev. D",
    volume = "35",
    pages = "1081",
    year = "1987"
}

@article{TASSO:1988,
    author = "{Tasso Collaboration}",
    collaboration = "TASSO",
    title = "{A search for particles with magnetic charge produced in $e^+ e^-$ annihilations at $\sqrt{s}=35$ GeV}",
    doi = "10.1007/BF01624358",
    journal = "Z. Phys. C",
    volume = "38",
    pages = "543",
    year = "1988"
}

@article{TRISTAN:1988,
    author = "Kinoshita, K. and Fujii, M. and Nakajima, K. and Price, P. B. and Tasaka, S.",
    title = "{Search for highly ionizing particles in $e^+ e^-$ annihilations at $\sqrt{s}=50$-$52$ GeV}",
    doi = "10.1103/PhysRevLett.60.1610",
    journal = "Phys. Rev. Lett.",
    volume = "60",
    pages = "1610",
    year = "1988"
}

@article{TRISTAN:1989,
    author = "Kinoshita, K. and Fujii, M. and Nakajima, K. and Price, P. B. and Tasaka, S.",
    title = "{Search for highly ionizing particles in $e^+ e^-$ annihilations at $\sqrt{s}=50$-$60.8$ GeV}",
    doi = "10.1016/0370-2693(89)90990-8",
    journal = "Phys. Lett. B",
    volume = "228",
    pages = "543",
    year = "1989"
}

@article{OPAL:2008,
    author = "{OPAL Collaboration}", 
    collaboration = "OPAL Collaboration",
    title = "{Search for Dirac magnetic monopoles in $e^+e^-$ collisions with the OPAL detector at LEP2}",
    eprint = "0707.0404",
    archivePrefix = "arXiv",
    primaryClass = "hep-ex",
    doi = "10.1016/j.physletb.2008.03.057",
    journal = "Phys. Lett. B",
    volume = "663",
    pages = "37",
    year = "2008"
}

@article{MODAL:1992,
    author = "Kinoshita, K. and Du, R. and Giacomelli, G. and Patrizii, L. and Predieri, F. and Serra, P. and Spurio, M. and Pinfold, J. L.",
    title = "{Search for highly ionizing particles in $e^+ e^-$ annihilations at $\sqrt{s}=91.1$ GeV}",
    doi = "10.1103/PhysRevD.46.R881",
    journal = "Phys. Rev. D",
    volume = "46",
    pages = "R881",
    year = "1992"
}

@article{L6-MODAL:1993,
    author = "Pinfold, J. L. and Du, R. and Kinoshita, K. and Lorazo, B. and Regimbald, M. and Price, B.",
    title = "{A Search for highly ionizing particles produced at the OPAL intersection point at LEP}",
    doi = "10.1016/0370-2693(93)90346-J",
    journal = "Phys. Lett. B",
    volume = "316",
    pages = "407",
    year = "1993"
}

@article{H1:2004,
    author = "{H1 Collaboration}",
    collaboration = "H1 Collaboration",
    title = "{A direct search for stable magnetic monopoles produced in positron-proton collisions at HERA}",
    eprint = "hep-ex/0501039",
    archivePrefix = "arXiv",
    reportNumber = "DESY-04-240",
    doi = "10.1140/epjc/s2005-02201-6",
    journal = "Eur. Phys. J. C",
    volume = "41",
    pages = "133",
    year = "2005"
}

@article{CDF:2005,
    author = "{CDF Collaboration}",
    collaboration = "CDF Collaboration",
    title = "{Direct Search for Dirac Magnetic Monopoles in $p\bar{p}$ Collisions at $\sqrt{s} = 1.96$ TeV}",
    eprint = "hep-ex/0509015",
    archivePrefix = "arXiv",
    doi = "10.1103/PhysRevLett.96.201801",
    journal = "Phys. Rev. Lett.",
    volume = "96",
    pages = "201801",
    year = "2006"
}

@article{Kalbfleisch:2003,
    author = "Kalbfleisch, G. R. and Luo, W. and Milton, K. A. and Smith, E. H. and Strauss, Michael G.",
    title = "{Limits on production of magnetic monopoles utilizing samples from the D0 and CDF detectors at the Fermilab Tevatron}",
    eprint = "hep-ex/0306045",
    archivePrefix = "arXiv",
    doi = "10.1103/PhysRevD.69.052002",
    journal = "Phys. Rev. D",
    volume = "69",
    pages = "052002",
    year = "2004"
}

@article{Kalbfleisch:2000,
    author = "Kalbfleisch, G. R. and Milton, K. A. and Strauss, Michael G. and Gamberg, L. P. and Smith, E. H. and Luo, W.",
    title = "{Improved Experimental Limits on the Production of Magnetic Monopoles}",
    eprint = "hep-ex/0005005",
    archivePrefix = "arXiv",
    doi = "10.1103/PhysRevLett.85.5292",
    journal = "Phys. Rev. Lett.",
    volume = "85",
    pages = "5292",
    year = "2000"
}

@article{D0:1998,
  author = "{D0 Collaboration}",
  title = {Search for Heavy Pointlike Dirac Monopoles},
  collaboration = {D0 Collaboration},
  journal = {Phys. Rev. Lett.},
  volume = {81},
  pages = {524},
  year = {1998},
  doi = {10.1103/PhysRevLett.81.524},
  eprint = "hep-ex/9803023",
  archivePrefix = "arXiv",
}

@article{Price:1987,
  title = {Search for Highly Ionizing Particles at the Fermilab Proton-Antiproton Collider},
  author = {Price, P. B. and Guoxiao, R. and Kinoshita, K.},
  journal = {Phys. Rev. Lett.},
  volume = {59},
  pages = {2523},
  year = {1987},
  doi = "10.1103/PhysRevLett.59.2523"
}

@article{Price:1990,
    author = "Price, P. B. and Guiru, J. and Kinoshita, K.",
    title = "{High-luminosity search for highly ionizing particles at the Fermilab collider}",
    doi = "10.1103/PhysRevLett.65.149",
    journal = "Phys. Rev. Lett.",
    volume = "65",
    pages = "149",
    year = "1990"
}

@article{Bertani:1990,
    author = "Bertani, M. and others",
    title = "{Search for Magnetic Monopoles at the Tevatron Collider}",
    doi = "10.1209/0295-5075/12/7/007",
    journal = "EPL",
    volume = "12",
    pages = "613",
    year = "1990"
}

@article{Hoffmann:1978,
    author = "Hoffmann, H. and Kantardjian, G. and Di Liberto, S. and Meddi, F. and Romano, G. and Rosa, G.",
    title = "{A new search for magnetic monopoles at the CERN-ISR with plastic detectors}",
    doi = "10.1007/BF02776247",
    journal = "Lett. Nuovo Cim.",
    volume = "23",
    pages = "357",
    year = "1978"
}

@article{Carrigan:1977,
    author = "Carrigan, Jr., Richard A. and Strauss, B. P. and Giacomelli, G.",
    title = "{Search for magnetic monopoles at the CERN Intersecting Storage Rings}",
    doi = "10.1103/PhysRevD.17.1754",
    journal = "Phys. Rev. D",
    volume = "17",
    pages = "1754",
    year = "1978"
}

@article{Aubert:1982,
    author = "Aubert, B. and Musset, P. and Price, M. and Vialle, J. P.",
    title = "{Search for magnetic monopoles in proton-antiproton interactions at 540~GeV cm energy}",
    doi = "10.1016/0370-2693(83)90488-4",
    journal = "Phys. Lett. B",
    volume = "120",
    pages = "465",
    year = "1983"
}

@article{Carrigan:1973,
    author = "Carrigan, R. A. and Nezrick, F. A. and Strauss, B. P.",
    title = "{Search for Magnetic-Monopole Production by 300-GeV Protons}",
    doi = "10.1103/PhysRevD.8.3717",
    journal = "Phys. Rev. D",
    volume = "8",
    pages = "3717",
    year = "1973"
}

@article{Carrigan:1974,
  title = "{Extension of Fermi National Accelerator Laboratory magnetic-monopole search to 400 GeV}",
  author = "Carrigan, R. A. and Nezrick, F. A. and Strauss, B. P.",
  journal = "Phys. Rev. D",
  volume = "10",
  pages = "3867",
  year = "1974",
  doi = "10.1103/PhysRevD.10.3867"
}

@article{PhysRevLett.79.3134,
  author = "He, Y. D.",
  title = "{Search for a Dirac Magnetic Monopole in High Energy Nucleus-Nucleus Collisions}",
  journal = "Phys. Rev. Lett.",
  volume = "79",
  pages = "3134",
  year = "1997",
  doi = "10.1103/PhysRevLett.79.3134",
}

@article{MoEDAL2016a,
      author         = "{MoEDAL Collaboration}",
title = "{Search for magnetic monopoles with the MoEDAL prototype trapping detector in 8 TeV proton-proton collisions at the LHC}",
    eprint = "1604.06645",
    archivePrefix = "arXiv",
    primaryClass = "hep-ex",
    reportNumber = "CERN-EP-2016-101",
    doi = "10.1007/JHEP08(2016)067",
    journal = "JHEP",
    volume = "08",
    pages = "067",
    year = "2016"
}

@article{MoEDAL2017a,
      author         = "{MoEDAL Collaboration}",
      title          = "{Search for Magnetic Monopoles with the MoEDAL Forward Trapping Detector in 13~TeV Proton-Proton Collisions at the LHC}",
      journal        = "Phys. Rev. Lett.",
      volume         = "118",
      year           = "2017",
      pages          = "061801",
      doi            = "10.1103/PhysRevLett.118.061801",
      archivePrefix  = "arXiv",
      eprint         = "1611.06817",
      primaryClass   = "hep-ex",
}

@article{MoEDAL2018,
   author = "{MoEDAL Collaboration}",
title = "{Search for magnetic monopoles with the MoEDAL forward trapping detector in 2.11 fb$^{-1}$ of 13 TeV proton-proton collisions at the LHC}",
    eprint = "1712.09849",
    archivePrefix = "arXiv",
    primaryClass = "hep-ex",
    reportNumber = "CERN-EP-2017-342",
    doi = "10.1016/j.physletb.2018.05.069",
    journal = "Phys. Lett. B",
    volume = "782",
    pages = "510--516",
    year = "2018"
}

@article{MoEDAL2019,
  author  = "{MoEDAL Collaboration}",
  title   = "{Magnetic Monopole Search with the Full MoEDAL Trapping Detector in 13 TeV $pp$ Collisions Interpreted in Photon-Fusion and Drell-Yan Production}",
  journal = "Phys. Rev. Lett.",
  volume  = "123",
  pages   = "021802",
  year    = "2019",
  doi     = "10.1103/PhysRevLett.123.021802",
  archivePrefix = "arXiv",
  eprint  = "1903.08491",
  primaryClass = "hep-ex",
}

@article{HOOFT1974276,
    author = "'t Hooft, Gerard",
    title = "{Magnetic Monopoles in Unified Gauge Theories}",
    reportNumber = "CERN-TH-1876",
    doi = "10.1016/0550-3213(74)90486-6",
    journal = "Nucl. Phys. B",
    volume = "79",
    pages = "276--284",
    year = "1974"
}

@article{Mavromatos:2020gwk,
    author = "Mavromatos, Nick E. and Mitsou, Vasiliki A.",
    title = "{Magnetic monopoles revisited: Models and searches at colliders and in the Cosmos}",
    eprint = "2005.05100",
    archivePrefix = "arXiv",
    primaryClass = "hep-ph",
    reportNumber = "KCL-PH-TH/2019-94, IFIC/20-12",
    doi = "10.1142/S0217751X20300124",
    journal = "Int. J. Mod. Phys. A",
    volume = "35",
    number = "23",
    pages = "2030012",
    year = "2020"
}

@article{Ahlen:1980xr,
    author = "Ahlen, S. P.",
    title = "{Theoretical and experimental aspects of the energy loss of relativistic heavily ionizing particles}",
    doi = "10.1103/RevModPhys.52.121",
    journal = "Rev. Mod. Phys.",
    volume = "52",
    pages = "121--173",
    year = "1980",
    related        = "Ahlen:1980xr-err", 
    relatedstring  = "Erratum:", 
}

@article{Harland-Lang:2023ohq,
    author = "Harland-Lang, L. A.",
    title = "{Exciting ions: A systematic treatment of ultraperipheral heavy ion collisions with nuclear breakup}",
    eprint = "2303.04826",
    archivePrefix = "arXiv",
    primaryClass = "hep-ph",
    doi = "10.1103/PhysRevD.107.093004",
    journal = "Phys. Rev. D",
    volume = "107",
    number = "9",
    pages = "093004",
    year = "2023"
}

@article{MoEDAL:2021vix,
    author = "{MoEDAL Collaboration}",
    collaboration = "MoEDAL",
    title = "{Search for magnetic monopoles produced via the Schwinger mechanism}",
    eprint = "2106.11933",
    archivePrefix = "arXiv",
    primaryClass = "hep-ex",
    doi = "10.1038/s41586-021-04298-1",
    journal = "Nature",
    volume = "602",
    number = "7895",
    pages = "63--67",
    year = "2022"
}

@article{MoEDAL:2024wbc,
    author = "{MoEDAL Collaboration}",
    collaboration = "MoEDAL",
    title = "{MoEDAL Search in the CMS Beam Pipe for Magnetic Monopoles Produced via the Schwinger Effect}",
    eprint = "2402.15682",
    archivePrefix = "arXiv",
    primaryClass = "nucl-ex",
    doi = "10.1103/PhysRevLett.133.071803",
    journal = "Phys. Rev. Lett.",
    volume = "133",
    number = "7",
    pages = "071803",
    year = "2024"
}

@article{Gould:2019myj,
    author = "Gould, Oliver and Ho, D. L.-J. and Rajantie, Arttu",
    title = "{Towards Schwinger production of magnetic monopoles in heavy-ion collisions}",
    eprint = "1902.04388",
    archivePrefix = "arXiv",
    primaryClass = "hep-th",
    reportNumber = "IMPERIAL-TP-2019-DH-01, HIP-2019-2/TH, IMPERIAL-TP-2019-DH-01; HIP-2019-2/TH",
    doi = "10.1103/PhysRevD.100.015041",
    journal = "Phys. Rev. D",
    volume = "100",
    number = "1",
    pages = "015041",
    year = "2019"
}

@article{Gould:2021bre,
    author = "Gould, Oliver and Ho, D. L.-J. and Rajantie, Arttu",
    title = "{Schwinger pair production of magnetic monopoles: Momentum distribution for heavy-ion collisions}",
    eprint = "2103.14454",
    archivePrefix = "arXiv",
    primaryClass = "hep-ph",
    reportNumber = "HIP-2021-12/TH, IMPERIAL-TP-2021-DH-05",
    doi = "10.1103/PhysRevD.104.015033",
    journal = "Phys. Rev. D",
    volume = "104",
    number = "1",
    pages = "015033",
    year = "2021"
}

@article{Drukier:1981fq,
    author = "Drukier, A. K. and Nussinov, S.",
    title = "{Monopole Pair Creation in Energetic Collisions: Is It Possible?}",
    reportNumber = "MPI-PAE/PTh 67/81",
    doi = "10.1103/PhysRevLett.49.102",
    journal = "Phys. Rev. Lett.",
    volume = "49",
    pages = "102",
    year = "1982"
}

@article{Klein:2016yzr,
    author = "Klein, S. R. and Nystrand, Joakim and Seger, Janet and Gorbunov, Yuri and Butterworth, Joey",
    title = "{STARlight: A Monte Carlo simulation program for ultra-peripheral collisions of relativistic ions}",
    eprint = "1607.03838",
    archivePrefix = "arXiv",
    primaryClass = "hep-ph",
    doi = "10.1016/j.cpc.2016.10.016",
    journal = "Comput. Phys. Commun.",
    volume = "212",
    pages = "258--268",
    year = "2017"
}

@article{Klein:2020fmr,
author = {Klein, Spencer and Steinberg, Peter},
title = {Photonuclear and Two-Photon Interactions at High-Energy Nuclear Colliders},
journal = {Annu. Rev. Nucl. Part. Sci.},
volume = {70},
number = {1},
pages = {323-354},
year = {2020},
doi = {10.1146/annurev-nucl-030320-033923},
    eprint = "2005.01872",
    archivePrefix = "arXiv",
    primaryClass = "nucl-ex",
}

@Booklet{Rimoldi:1097789,
      author         = "{ATLAS Collaboration}",
      title         = "{First Report of the Simulation Optimization Group}",
      year          = "2008",
      url           = "https://cds.cern.ch/record/1097789",
      howpublished   = "{ATL-SOFT-PUB-2008-002}",
}

@article{Agostinelli:2002hh,
      author         = "Agostinelli, S. and others",
      title          = "{\textsc{Geant4} -- a simulation toolkit}",
      journal        = "Nucl. Instrum. Meth. A",
      volume         = "506",
      year           = "2003",
      pages          = "250",
      doi            = "10.1016/S0168-9002(03)01368-8",
      reportNumber   = "SLAC-PUB-9350, FERMILAB-PUB-03-339",
      SLACcitation   = "%%CITATION = NUIMA,A506,250;%%"
}

@Article{Read:2002hq,
    Author = {Read, Alexander L.},
    Journal = {J. Phys. G},
    Pages = {2693},
    doi = "10.1088/0954-3899/28/10/313",
    Title = "{Presentation of search results: the \(CL_S\) technique}",
    Volume = {28},
    Year = {2002}
}

@Article{PERF-2007-01,
    author         = "{ATLAS Collaboration}",
    title          = "{The ATLAS Experiment at the CERN Large Hadron Collider}",
    journal        = "JINST",
    volume         = "3",
    year           = "2008",
    pages          = "S08003",
    doi            = "10.1088/1748-0221/3/08/S08003",
    primaryClass   = "hep-ex",
}

@Article{SOFT-2010-01,
    author         = "{ATLAS Collaboration}",
    title          = "{The ATLAS Simulation Infrastructure}",
    journal        = "Eur. Phys. J. C",
    volume         = "70",
    year           = "2010",
    pages          = "823",
    doi            = "10.1140/epjc/s10052-010-1429-9",
    eprint         = "1005.4568",
    archivePrefix  = "arXiv",
    primaryClass   = "physics.ins-det",
}

@Article{EXOT-2011-27,
    author         = "{ATLAS Collaboration}",
    title          = "{Search for Magnetic Monopoles in \(\sqrt{s} = 7\,\text{TeV}\) \(pp\) Collisions with the ATLAS Detector}",
    journal        = "Phys. Rev. Lett.",
    volume         = "109",
    year           = "2012",
    pages          = "261803",
    doi            = "10.1103/PhysRevLett.109.261803",
    reportNumber   = "CERN-PH-EP-2012-177",
    eprint         = "1207.6411",
    archivePrefix  = "arXiv",
    primaryClass   = "hep-ex",
}

@Article{STDM-2011-01,
    author         = "{ATLAS Collaboration}",
    title          = "{Rapidity gap cross sections measured with the ATLAS detector in \(pp\) collisions at \(\sqrt{s} = 7\,\text{TeV}\)}",
    journal        = "Eur. Phys. J. C",
    volume         = "72",
    year           = "2012",
    pages          = "1926",
    doi            = "10.1140/epjc/s10052-012-1926-0",
    reportNumber   = "CERN-PH-EP-2011-220",
    eprint         = "1201.2808",
    archivePrefix  = "arXiv",
    primaryClass   = "hep-ex",
}

@Article{DAPR-2012-01,
    author         = "{ATLAS Collaboration}",
    title          = "{Characterisation and mitigation of beam-induced backgrounds observed in the ATLAS detector during the 2011 proton--proton run}",
    journal        = "JINST",
    volume         = "8",
    year           = "2013",
    pages          = "P07004",
    doi            = "10.1088/1748-0221/8/07/P07004",
    reportNumber   = "CERN-PH-EP-2012-313",
    eprint         = "1303.0223",
    archivePrefix  = "arXiv",
    primaryClass   = "hep-ex",
}

@Article{PERF-2012-05,
    author         = "{ATLAS Collaboration}",
    title          = "{A neural network clustering algorithm for the ATLAS silicon pixel detector}",
    journal        = "JINST",
    volume         = "9",
    year           = "2014",
    pages          = "P09009",
    doi            = "10.1088/1748-0221/9/09/P09009",
    reportNumber   = "CERN-PH-EP-2014-120",
    eprint         = "1406.7690",
    archivePrefix  = "arXiv",
    primaryClass   = "hep-ex",
}

@Article{DAPR-2014-01,
    author         = "{ATLAS Collaboration}",
    title          = "{Beam-induced and cosmic-ray backgrounds observed in the ATLAS detector during the LHC 2012 proton--proton running period}",
    journal        = "JINST",
    volume         = "11",
    year           = "2016",
    pages          = "P05013",
    doi            = "10.1088/1748-0221/11/05/P05013",
    reportNumber   = "CERN-EP-2016-029",
    eprint         = "1603.09202",
    archivePrefix  = "arXiv",
    primaryClass   = "hep-ex",
}

@Article{EXOT-2014-16,
    author         = "{ATLAS Collaboration}",
    title          = "{Search for magnetic monopoles and stable particles with high electric charges in \(8\,\text{TeV}\) \(pp\) collisions with the ATLAS detector}",
    journal        = "Phys. Rev. D",
    volume         = "93",
    year           = "2016",
    pages          = "052009",
    doi            = "10.1103/PhysRevD.93.052009",
    reportNumber   = "CERN-PH-EP-2015-174",
    eprint         = "1509.08059",
    archivePrefix  = "arXiv",
    primaryClass   = "hep-ex",
}

@Article{PERF-2014-07,
    author         = "{ATLAS Collaboration}",
    title          = "{Topological cell clustering in the ATLAS calorimeters and its performance in LHC Run~1}",
    journal        = "Eur. Phys. J. C",
    volume         = "77",
    year           = "2017",
    pages          = "490",
    doi            = "10.1140/epjc/s10052-017-5004-5",
    reportNumber   = "CERN-PH-EP-2015-304",
    eprint         = "1603.02934",
    archivePrefix  = "arXiv",
    primaryClass   = "hep-ex",
}

@Article{PERF-2015-07,
    author         = "{ATLAS Collaboration}",
    title          = "{Study of the material of the ATLAS inner detector for Run~2 of the LHC}",
    journal        = "JINST",
    volume         = "12",
    year           = "2017",
    pages          = "P12009",
    doi            = "10.1088/1748-0221/12/12/P12009",
    reportNumber   = "CERN-EP-2017-081",
    eprint         = "1707.02826",
    archivePrefix  = "arXiv",
    primaryClass   = "hep-ex",
}

@Article{HION-2016-02,
    author         = "{ATLAS Collaboration}",
    title          = "{Exclusive dimuon production in ultraperipheral Pb+Pb collisions at \(\sqrt{s_{\text{NN}}} = 5.02\,\text{TeV}\) with ATLAS}",
    journal        = "Phys. Rev. C",
    volume         = "104",
    year           = "2021",
    pages          = "024906",
    doi            = "10.1103/PhysRevC.104.024906",
    reportNumber   = "CERN-EP-2020-138",
    eprint         = "2011.12211",
    archivePrefix  = "arXiv",
    primaryClass   = "hep-ex",
}

@Article{HION-2016-06,
    author         = "{ATLAS Collaboration}",
    title          = "{Measurement of the azimuthal anisotropy of charged particles produced in \(\sqrt{s_{\text{NN}}} = 5.02\,\text{TeV}\) Pb+Pb collisions with the ATLAS detector}",
    journal        = "Eur. Phys. J. C",
    volume         = "78",
    year           = "2018",
    pages          = "997",
    doi            = "10.1140/epjc/s10052-018-6468-7",
    reportNumber   = "CERN-EP-2018-194",
    eprint         = "1808.03951",
    archivePrefix  = "arXiv",
    primaryClass   = "hep-ex",
}

@Article{DAPR-2017-01,
    author         = "{ATLAS Collaboration}",
    title          = "{Comparison between simulated and observed LHC beam backgrounds in the ATLAS experiment at \({E_{\text{beam}}}\ = 4\,\text{TeV}\)}",
    journal        = "JINST",
    volume         = "13",
    year           = "2018",
    pages          = "P12006",
    doi            = "10.1088/1748-0221/13/12/P12006",
    reportNumber   = "CERN-EP-2018-240",
    eprint         = "1810.04450",
    archivePrefix  = "arXiv",
    primaryClass   = "hep-ex",
}

@Article{EXOT-2017-20,
    author         = "{ATLAS Collaboration}",
    title          = "{Search for Magnetic Monopoles and Stable High-Electric-Charge Objects in \(13\,\text{TeV}\) Proton--Proton Collisions with the ATLAS Detector}",
    journal        = "Phys. Rev. Lett.",
    volume         = "124",
    year           = "2020",
    pages          = "031802",
    doi            = "10.1103/PhysRevLett.124.031802",
    reportNumber   = "CERN-EP-2019-084",
    eprint         = "1905.10130",
    archivePrefix  = "arXiv",
    primaryClass   = "hep-ex",
}

@Article{STDM-2017-28,
    author         = "{ATLAS Collaboration}",
    title          = "{Measurement of distributions sensitive to the underlying event in inclusive \(Z\) boson production in \(pp\) collisions at \(\sqrt{s} = 13\,\text{TeV}\) with the ATLAS detector}",
    journal        = "Eur. Phys. J. C",
    volume         = "79",
    year           = "2019",
    pages          = "666",
    doi            = "10.1140/epjc/s10052-019-7162-0",
    reportNumber   = "CERN-EP-2019-064",
    eprint         = "1905.09752",
    archivePrefix  = "arXiv",
    primaryClass   = "hep-ex",
}

@Article{HION-2018-27,
    author         = "{ATLAS Collaboration}",
    title          = "{Two-particle azimuthal correlations in photonuclear ultraperipheral Pb+Pb collisions at \(5.02\,\text{TeV}\) with ATLAS}",
    journal        = "Phys. Rev. C",
    volume         = "104",
    year           = "2021",
    pages          = "014903",
    doi            = "10.1103/PhysRevC.104.014903",
    reportNumber   = "CERN-EP-2020-246",
    eprint         = "2101.10771",
    archivePrefix  = "arXiv",
    primaryClass   = "hep-ex",
}

@Article{PIX-2018-001,
    author         = "Abbott, B. and others",
    title          = "{Production and integration of the ATLAS Insertable B-Layer}",
    journal        = "JINST",
    volume         = "13",
    year           = "2018",
    pages          = "T05008",
    doi            = "10.1088/1748-0221/13/05/T05008",
    eprint         = "1803.00844",
    archivePrefix  = "arXiv",
    primaryClass   = "physics.ins-det",
}

@Article{EXOT-2019-33,
    author         = "{ATLAS Collaboration}",
    title          = "{Search for magnetic monopoles and stable particles with high electric charges in \(\sqrt{s} = 13\,\text{TeV}\) \(pp\) collisions with the ATLAS detector}",
    journal        = "JHEP",
    volume         = "11",
    year           = "2023",
    pages          = "112",
    doi            = "10.1007/JHEP11(2023)112",
    reportNumber   = "CERN-EP-2023-150",
    eprint         = "2308.04835",
    archivePrefix  = "arXiv",
    primaryClass   = "hep-ex",
}

@Article{DAPR-2021-01,
    author         = "{ATLAS Collaboration}",
    title          = "{Luminosity determination in \(pp\) collisions at \(\sqrt{s} = 13\,\text{TeV}\) using the ATLAS detector at the LHC}",
    journal        = "Eur. Phys. J. C",
    volume         = "83",
    year           = "2023",
    pages          = "982",
    doi            = "10.1140/epjc/s10052-023-11747-w",
    reportNumber   = "CERN-EP-2022-281",
    eprint         = "2212.09379",
    archivePrefix  = "arXiv",
    primaryClass   = "hep-ex",
}

@Article{HION-2021-16,
    author         = "{ATLAS Collaboration}",
    title          = "{Exclusive dielectron production in ultraperipheral Pb+Pb collisions at \(\sqrt{s_{\text{NN}}} = 5.02\,\text{TeV}\) with ATLAS}",
    journal        = "JHEP",
    volume         = "06",
    year           = "2023",
    pages          = "182",
    doi            = "10.1007/JHEP06(2023)182",
    reportNumber   = "CERN-EP-2022-130",
    eprint         = "2207.12781",
    archivePrefix  = "arXiv",
    primaryClass   = "nucl-ex",
}

@Article{SOFT-2022-02,
    author         = "{ATLAS Collaboration}",
    title          = "{Software and computing for Run~3 of the ATLAS experiment at the LHC}",
    year           = "2024",
    reportNumber   = "CERN-EP-2024-100",
    eprint         = "2404.06335",
    archivePrefix  = "arXiv",
    primaryClass   = "hep-ex",
}

@Article{HION-2023-01,
    author         = "{ATLAS Collaboration}",
    title          = "{Search for magnetic monopole pair production in ultraperipheral Pb+Pb collisions at \(\sqrt{s_{\text{NN}}} = 5.36\,\text{TeV}\) with the ATLAS detector at the LHC}",
    year           = "2024",
    reportNumber   = "CERN-EP-2024-209",
    eprint         = "2408.11035",
    archivePrefix  = "arXiv",
    primaryClass   = "nucl-ex",
}

@Booklet{ATL-SOFT-PUB-2023-001,
    author         = "{ATLAS Collaboration}",
    title          = "{ATLAS Computing Acknowledgements}",
    howpublished   = "{ATL-SOFT-PUB-2023-001}",
    url            = "https://cds.cern.ch/record/2869272",
    year           = "2023",
}

@Report{ATLAS-TDR-2010-19,
    author         = "{ATLAS Collaboration}",
    title          = "{ATLAS Insertable B-Layer Technical Design Report}",
    type           = "ATLAS-TDR-19; CERN-LHCC-2010-013",
    year           = "2010",
    url            = "https://cds.cern.ch/record/1291633",
    related        = "ATLAS-TDR-2010-19-addm",
    relatedstring  = "Addendum:",
}
\clearpage
 
\begin{flushleft}
\hypersetup{urlcolor=black}
{\Large The ATLAS Collaboration}

\bigskip

\AtlasOrcid[0000-0002-6665-4934]{G.~Aad}$^\textrm{\scriptsize 105}$,
\AtlasOrcid[0000-0001-7616-1554]{E.~Aakvaag}$^\textrm{\scriptsize 17}$,
\AtlasOrcid[0000-0002-5888-2734]{B.~Abbott}$^\textrm{\scriptsize 124}$,
\AtlasOrcid[0000-0002-0287-5869]{S.~Abdelhameed}$^\textrm{\scriptsize 120a}$,
\AtlasOrcid[0000-0002-1002-1652]{K.~Abeling}$^\textrm{\scriptsize 57}$,
\AtlasOrcid[0000-0001-5763-2760]{N.J.~Abicht}$^\textrm{\scriptsize 51}$,
\AtlasOrcid[0000-0002-8496-9294]{S.H.~Abidi}$^\textrm{\scriptsize 30}$,
\AtlasOrcid[0009-0003-6578-220X]{M.~Aboelela}$^\textrm{\scriptsize 46}$,
\AtlasOrcid[0000-0002-9987-2292]{A.~Aboulhorma}$^\textrm{\scriptsize 36e}$,
\AtlasOrcid[0000-0001-5329-6640]{H.~Abramowicz}$^\textrm{\scriptsize 156}$
% ...
\bigskip
\\

% ...
$^{17}$Department for Physics and Technology, University of Bergen, Bergen; Norway.\\
% ...
$^{30}$Physics Department, Brookhaven National Laboratory, Upton NY; United States of America.\\
% ...
$^{36}$$^{(a)}$Facult\'e des Sciences Ain Chock, Universit\'e Hassan II de Casablanca;$^{(b)}$Facult\'{e} des Sciences, Universit\'{e} Ibn-Tofail, K\'{e}nitra;$^{(c)}$Facult\'e des Sciences Semlalia, Universit\'e Cadi Ayyad, LPHEA-Marrakech;$^{(d)}$LPMR, Facult\'e des Sciences, Universit\'e Mohamed Premier, Oujda;$^{(e)}$Facult\'e des sciences, Universit\'e Mohammed V, Rabat;$^{(f)}$Institute of Applied Physics, Mohammed VI Polytechnic University, Ben Guerir; Morocco.\\
% ...
$^{46}$Physics Department, Southern Methodist University, Dallas TX; United States of America.\\
% ...
$^{51}$Fakult\"{a}t Physik , Technische Universit{\"a}t Dortmund, Dortmund; Germany.\\
% ...
$^{57}$II. Physikalisches Institut, Georg-August-Universit\"{a}t G\"ottingen, G\"ottingen; Germany.\\
% ...
$^{105}$CPPM, Aix-Marseille Universit\'e, CNRS/IN2P3, Marseille; France.\\
% ...
$^{120}$$^{(a)}$New York University Abu Dhabi, Abu Dhabi;$^{(b)}$United Arab Emirates University, Al Ain; United Arab Emirates.\\
% ...
$^{124}$Homer L. Dodge Department of Physics and Astronomy, University of Oklahoma, Norman OK; United States of America.\\
% ...
$^{156}$Raymond and Beverly Sackler School of Physics and Astronomy, Tel Aviv University, Tel Aviv; Israel.\\
% ...
$^{a}$ Also Affiliated with an institute covered by a cooperation agreement with CERN.\\
$^{b}$ Also at An-Najah National University, Nablus; Palestine.\\
$^{c}$ Also at Borough of Manhattan Community College, City University of New York, New York NY; United States of America.\\
$^{d}$ Also at Center for High Energy Physics, Peking University; China.\\
$^{e}$ Also at Center for Interdisciplinary Research and Innovation (CIRI-AUTH), Thessaloniki; Greece.\\
$^{f}$ Also at CERN, Geneva; Switzerland.\\
% ...

\end{flushleft}


%

%
%

%

%
%
%
%

\end{document}